# Indexes in Microsoft SQL Server

Sourav Mukherjee

Senior Database Administrator &

PhD student at University of the Cumberlands

Chicago, United States

**Abstract**

Indexes are the best apposite choice for quickly retrieving the records. This is nothing but cutting down the number of Disk IO. Instead of scanning the complete table for the results, we can decrease the number of IO's or page fetches using index structures such as B-Trees or Hash Indexes to retrieve the data faster. The most convenient way to consider an index is to think like a dictionary. It has words and its corresponding definitions against those words. The dictionary will have an index on "word" because when we open a dictionary and we want to fetch its corresponding word quickly, then find its definition. The dictionary generally contains just a single index - an index ordered by word. When we modify any record and change the corresponding value of an indexed column in a clustered index, the database might require moving the entire row into a separately new position to maintain the rows in the sorted order. This action is essentially turned into an update query into a DELETE followed by an INSERT, and it decreases the performance of the query. The clustered index in the table can often be available on the primary key or a foreign key column because key values usually do not modify once a record is injected into the database.

## Indexes in Microsoft SQL Server

The index is a structure in SQL Server either on-disk or in-memory structure associated with a table or View that is used to quickly identify rows or a specific set of rows from the table or views. We can imagine indexes like the front of the book with the name like index page that describes the primary key and the end of the book we have a glossary which talks about the non-clustered indexes. The index includes the keys put together from one or more columns in the table or the view. The keys are stored in the B-tree structure that allows SQL Server to find the row(s) associated with the key values fast and effectively. For indexes which are on-disk, the keys are stored in a structure (B-Tree) which allows SQL Server to extract the row or rows associated with the key values quickly and efficiently.

The B-Tree structure allows the SQL Server Engine to move faster while moving through the table rows based on the index keys either letting it navigate right or left and helps to retrieve the result directly rather scanning all the table records.

An index stores data in a table organized logically with rows and columns, and physically stored in a row-wise data format called rowstore or if the records are stored in a column-wise data format, known as columnstore.

An index is defined on one or more columns, called Key columns. The key columns are also known as the index key. The index is structured by the key columns. Often the indexes are formed with more than one key column which is classified as a composite index.

The ideal Index B-Tree structure looks like the diagram mentioned below. In general, the index has a root page with Zero/more intermediate levels followed by a leaf level. 1 page is an 8 KB chunk of the data file, with a header and footer and is identified by a combination of the File ID and Page number.

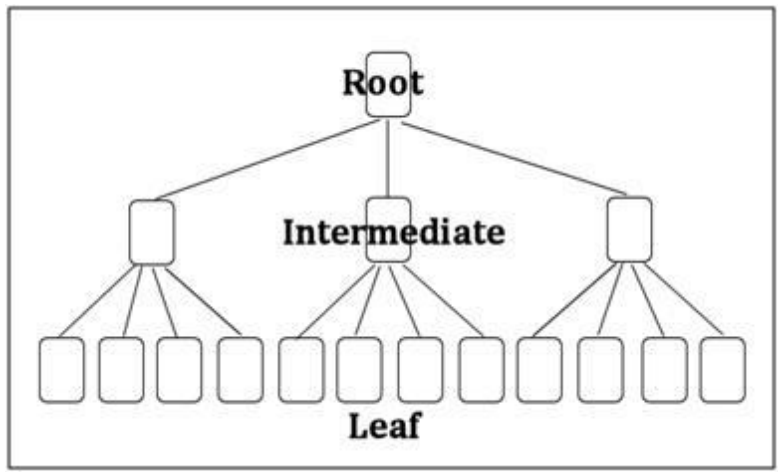

**Fig 1: Index structure (Root, Intermediate and Leaf structure) [8]**

Generally, the root page is at the top of the tree diagram and from here the SQL Server initiates the data search process. Next, the leaf level which is the bottom level of the nodes that contains the data pages we look for. The size of the index depends upon the count of the data stored in the leaf pages. At the middle it is the intermediate level, it is one or multiple levels between the root and the leaf levels which holds the index key values and its pointers to the next intermediate level pages or the leaf data pages. Generally, the number of intermediate levels be subject to the amount of data stored in the index. It is also referred to as the depth of the index. The efficiency of the index greatly depends on the depth of the index. The index structure figure #1 showcases that it has a depth of 3.

The B-Tree structure looks more like an Inverted Tree Structure.

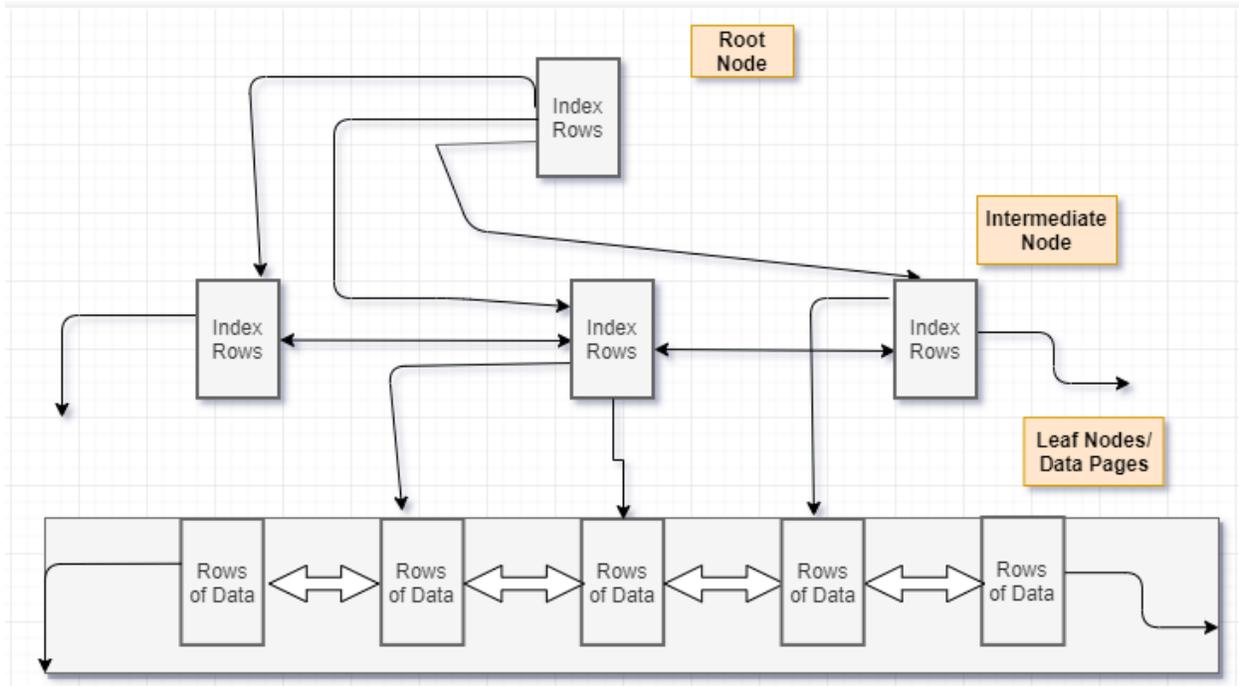

**Fig 2: B-Tree Structure**

Let's assume that we create an index in one of the database tables against one of the ID columns. Now, once we run any query to extract a set of rows from that table based on the ID value in the rows, SQL Server engine will start navigating the records from the root node to identify which page to reference in the top intermediate level and then it continues to go down through the other intermediate nodes to identify the address of the next intermediate node, until it reaches the target leaf node that eventually contains the requested data row or pointer to that row in the main table depends on the type of the index. If we create an index on the primary key column and search for a row of any data based on a primary key value, SQL Server first looks for that value in the index, and finally uses the index to rapidly find the entire row of data. Without the index, a table scan is inevitable in order to locate the row, which degrades the performance greatly.

In SQL Server, Indexes may have a relatively large number of nodes at each level. This is quite helpful for the index by avoiding the need for excessive depth within the index.

**Index Depth**: It signifies the number of levels from the index root node to the leaf nodes. An index that is very deep will suffer from performance degradation problems. In contrast, an index with many nodes at each level can produce a very flat index structure. It is very common to have an index with 4-5 levels.

There are few other key index measurements that control the index effectiveness.

**Index Density:** This property in the index is a measure of the deficiency of uniqueness of the data in a table. A dense column is defined as the one that has a high number of duplicates.

**Index Selectivity:** This property is a measure of how many rows scanned as compared to the total number of rows. An index with high selectivity means a trivial number of rows are scanned when related to the total number of rows.

A poorly designed index in the tables and a lack of appropriate indexes are the major cause of database application bottlenecks. Designing well-organized indexes are the key to achieving a good database and application performance.

Identifying the right index and its workload identification is a challenging job as it requires a complex balancing act between the query speed vs updated cost. If the index size is smaller due to fewer columns, that takes up less storage and maintenance overhead. On the other hand, the wider indexes cover more queries. The superior choice would be to experience with different index structure and designs before choosing the useful index. Indexes can be added, updated, modified and dropped online/offline without changing the database schema and/or the application design. Hence this is always a recommended practice to experiment with fewer and different indexes.

**Index Design**

The following tasks make up our recommended strategy for designing indexes:

1. Understand the characteristics of the database itself.
   - If the database is an OLTP kind, memory-optimized tables (for SQL 2014+) and the indexes are suitable as that offers latch-fee design.
   - If the database is an OLAP/Datawarehouse/Decision Support System kind, then it is more advisable to use a columnstore index (SQL Server 2012+).
2. Understanding the characteristics of the columns used in the queries.
   Example: If the columns in the table have integer data types and have nonnull or unique columns then an index is most suitable.
3. Identifying the characteristic of the most frequently used queries.
4. Example: Find out the joining statements between multiple columns and from there find out the best possible index to be used.
5. Identify periodically which index options might enhance performance when the index is created or maintained.
   Example: Use an ONLINE index option while creating a clustered index on an existing large table. The ONLINE option allows for concurrent activity on the underlying data to continue while the index is being created or rebuilt.
6. Identify the ideal storage location for the index.
   We can place the nonclustered index in the same filegroup as the underlying table or can be placed on a different filegroup. The index storage location may improve query performance by increasing disk I/O performance.

Example, storing a nonclustered index on a filegroup that reside is on a different disk than the table filegroup may improve query performance because multiple disks can be read at the same time.

**Index Design Guidance**

Creating the best indexes are a tough job. Understanding the characteristics of the database, queries and data columns can help to design the optimal indexes.

Below are the recommendations while designing an index:

- Many indexes on a table affect the performance of DML operations.
  **Example**: If a column is used in several indexes and someone executes the UPDATE statement, that modifies the column's data, each index that contains that column should be updated and the column in the underlying base table.
- Avoid over indexing heavily updated tables.
- Avoid indexing against smaller tables.
- Indexes on views can provide significant performance gains if the view contains table joins, aggregations or a combination of two.
- Check the queries and statements using Actual Vs Estimated execution plan and identify the areas of improvements.
- Choose the right fill factor. Through this setting, we can customize the initial storage characteristics of the index to optimize its performance or maintenance.

## Query Considerations

- Create nonclustered indexes on the columns that are frequently used in predicates and join conditions in queries.
- Covering indexes (especially for nonclustered indexes) can improve query performance because all the data needed to meet the requirements of the query exists within the index itself.
- Write queries that insert or modify as many rows as possible in a single statement, instead of using multiple queries to update the same rows.

## Column considerations

- Keep the length of the index key short for clustered indexes. Additionally, clustered indexes benefit from being created on unique or nonnull columns.
- Columns that are of the ntext, text, image, varchar(max), nvarchar(max), and varbinary(max) data types cannot be specified as index key columns.
- An xml data type can only be a key column only in an XML index.
- Examine column uniqueness. A unique index instead of a nonunique index on the same combination of columns provide additional information for the query optimizer that makes the index more useful.
- Examine data distribution in the column. Most dominantly, a long-running query is instigated by indexing a column with insufficient unique values, or by running a join on such a column. This is a foremost problem with the data and query, and generally cannot be resolute without categorizing this situation.

- A well-designed filtered index can improve query performance, reduce index maintenance costs, and reduce storage costs.
- Consider the order of the columns if the index will contain multiple columns. The column that is used in the WHERE clause in an equal to (=), greater than (>), less than (<), or BETWEEN search condition, or participates in a join, should be placed first. Additional columns should be ordered based on their level of distinctness, that is, from the most distinct to the least distinct.

There are different types of indexes which exist in Microsoft SQL Server.

- Clustered
- Nonclustered
- Unique
- Filtered
- Columnstore
- Hash
- Memory-Optimized Nonclustered

**Different kind of indexes**

- **Clustered Index**
    - Clustered indexes sort and store data rows in the table or view depending on their key values. These columns are included in the index definition. There may be just one clustered index in each table as the data rows can be stored in a single order.

- o   The only time when the data rows in a table are stored in sorted order is when the table contains a clustered index. When a table contains a clustered index, the table is called a clustered table. If a table does not contain any clustered index, its data rows are stored in an unordered structure called a heap.

- **Nonclustered Index**
    - o   Nonclustered indexes do possess a structure that is separated from the data rows. A nonclustered index covers the nonclustered index key values and each key-value pair has a pointer to the data row that contains the key value.
    - o   The pointer from an index row in a nonclustered index to a data row is called a row locator. The structure of the row locator hinge on whether the data pages are in a heap or a clustered table. For a heap, a row locator is generally a pointer to the row. For a clustered table, the row locator is called the clustered index key.
    - o   We can add nonkey columns to the leaf level of the nonclustered index. It by-passes existing index key limits, and perform fully covered, indexed, queries.

- **Unique Index**
    - o   A unique index guarantees that the index key contains no duplicate values and therefore every row in the table is in some way unique.
    - o   Specifying a unique index makes sense only when uniqueness is a characteristic of the data itself.

- Creating a PRIMARY KEY or UNIQUE constraint automatically creates a unique index on the specified columns. There are no noteworthy differences between creating a UNIQUE constraint and generating a unique index independent of a constraint.

The benefits of unique indexes include the following:

- Data integrity of the defined columns is ensured.
- Additional information helpful to the query optimizer is provided.

- **Filtered Indexes**
  - When a column only has a small number of relevant values for queries, we can create a filtered index on the subset of values. For example, when the values in a column are mostly NULL and the query selects only from the non-NULL values, we can create a filtered index for the non-NULL data rows. The subsequent index will be smaller and would cost less to maintain as compared to a full-table nonclustered index included on the same key columns.
  - When a table has heterogeneous data rows, we can create a filtered index for one or more categories of data.
  - The query optimizer can indicate a filtered index for the query irrespective of whether it does or does not cover the query. However, the query optimizer is more possible to choose a filtered index if it covers the query.
  - A column in the filtered index expression does not need to be a key or included column in the filtered index definition if the filtered index expression is corresponding to the query predicate and also the query does not give back the column in the filtered index expression with the query results.

- **Columnstore Index**

    o   A columnstore index is a technology for storing, retrieving and managing data by using a columnar data format, called a columnstore.

    o   A columnstore is data that is logically organized as a table with rows and columns, and physically stored in a column-wise data format.

    o   A rowstore is data that is logically organized as a table with rows and columns, and then physically stored in a row-wise data format. This is the traditional way of storing relational table data such as a heap or clustered B-tree index.

    o   The deltastore is a holding place for rows that are too few to be compressed into the columnstore. The deltastore stores the rows in rowstore format.

    o   An in-memory table can have one columnstore index. We can create it when the table is created or add it later with ALTER TABLE (Transact-SQL).

    o   The nonclustered columnstore index definition supports using a filtered condition. To minimize the performance impact of adding a columnstore index on an OLTP table, use a filtered condition to create a nonclustered columnstore index on only the cold data of our operational workload.

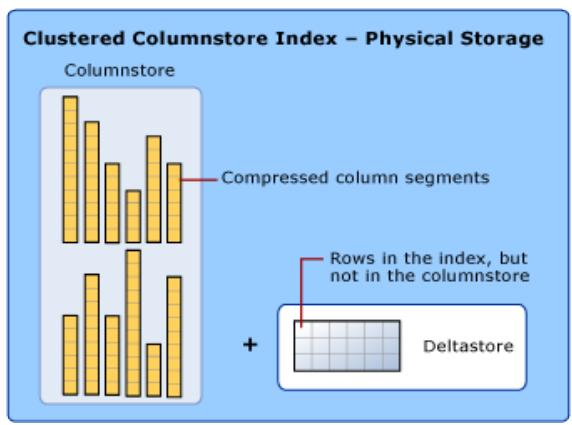

**Fig 2: Clustered Columnstore Index [1]**

- **Hash Index**
    - All memory-optimized tables must have at least one index because it is the indexes that connect the rows together. On a memory-optimized table, every index is also memory-optimized. Hash indexes are one of the possible index types in a memory-optimized table.
    - A hash index consists of an array of pointers, and each element of the array is called a hash bucket. Each bucket is of 8 bytes that are used to stock the memory address of a linked list of key entries.

The performance of a hash index is:

- Excellent when the predicate in the WHERE clause specifies an exact value for each column in the hash index key. A hash index will return to a scan assumed an inequality predicate.
- Poor when the predicate in the WHERE clause looks for a range of values in the index key.
- Poor when the predicate in the WHERE clause stipulates one specific value for the first column of a two-column hash index key but does not specify a value for other columns of the key.

- **Memory-Optimized Nonclustered Index**
    - Nonclustered indexes are one of the possible index types in a memory-optimized table. In-memory nonclustered indexes are implemented using a data structure called a Bw-

- Tree, originally envisioned and described by Microsoft Research in 2011. A Bw-Tree is a lock and latch-free difference of a B-Tree.
    - The performance of a nonclustered index is better than nonclustered hash indexes when querying a memory-optimized table with inequality predicates.

## Conclusions and Future Study

The query optimizer built with SQL Server selects the highly effective index in most cases. The overall index design strategy gives a variety of index choices for the query optimizer to identify the best one and we trust the component to do the right decision. This process decreases the overall analysis time and turns out to be a good performance over a variety of situations. To identify which indexes the query optimizer uses for a specific query, in SQL Server Management Studio, on the Query menu, select Include Actual Execution Plan and/or Estimated Execution Plan. It is also not a good practice to always try to equate the index usage with good performance. Sometimes it is also possible that an incorrect index choice may also cause less than the optimal performance. Therefore, the job of the query optimizer is to select an index, or grouping of indexes, only when it will improve performance, and to avoid indexed retrieval when it will hinder performance.

**AUTHOR'S PROFILE**

Sourav Mukherjee is a Senior Database Administrator and Data Architect based out of Chicago. He has more than 12 years of experience working with Microsoft SQL Server Database Platform. His work focusses in Microsoft SQL Server started with SQL Server 2000. Being a consultant architect, he has worked with different Chicago based clients. He has helped many companies in designing and maintaining their high availability solutions, developing and designing appropriate security models and providing query tuning guidelines to improve the overall SQL Server health, performance and simplifying the automation needs. He is passionate about SQL Server Database and the related community and contributing to articles in different SQL Server Public sites and Forums helping the community members. He holds a bachelor's degree in Computer Science & Engineering followed by a master's degree in Project

Management. Currently pursuing Ph.D. In Information Technology from the University of the Cumberlands. His areas of research interest include RDBMS, distributed database, Cloud Security, AI and Machine Learning. He is an MCT (Microsoft Certified Trainer) since 2017 and holds other premier certifications such as MCP, MCTS, MCDBA, MCITP, TOGAF, Prince2, Certified Scrum Master and ITIL.